\documentclass[conference]{IEEEtran}
\usepackage{caption,subcaption}

\RequirePackage{etex}
\usepackage{graphicx}
\usepackage{adjustbox}
\usepackage{color}
\usepackage{amsfonts}
\usepackage{float}
\usepackage{amsmath}
\usepackage{boldline}
\usepackage{amsthm}
\usepackage{epstopdf}
\usepackage{changepage}
\usepackage{latexsym}
\usepackage{tikz}
\usepackage{cases}
\usepackage{amssymb}
\usepackage{longtable}
\usepackage{supertabular}
\usepackage{tabu}
\usepackage{multirow}
\usepackage{multicol}
\usepackage{booktabs}
\usepackage{arydshln}
\usepackage{setspace}
\usepackage{algpseudocode}
\usepackage{etex}
\usepackage{algcompatible}
\usepackage{amsmath,amssymb,amsthm,mathrsfs,amsfonts,dsfont}
\usepackage{multicol}
\usepackage{booktabs}
\usepackage{arydshln}
\usepackage{setspace}
\usepackage{etex}
\usepackage{graphicx}
\usepackage{color}
\usepackage{amsfonts}
\usepackage{float}
\usepackage{amsmath}
\usepackage{amsthm}
\usepackage{epstopdf}
\usepackage{changepage}
\usepackage{latexsym}
\usepackage{tikz}
\usepackage{cases}
\usepackage{longtable}
\usepackage{supertabular}
\usepackage{tabu}
\usepackage{multirow}
\usepackage{multicol}
\usepackage{booktabs}
\usepackage{arydshln}
\usepackage{setspace}
\usepackage{amsthm}
\usepackage{graphicx}
\usepackage{color}
\usepackage{amsfonts}
\usepackage{float}
\usepackage{amsmath}
\usepackage{amsthm}
\usepackage{epstopdf}
\usepackage{changepage}
\usepackage{latexsym}
\usepackage{tikz}
\usepackage{cases}
\usepackage{longtable}
\usepackage{supertabular}
\usepackage{tabu}
\usepackage{multirow}
\usepackage{multicol}
\usepackage{booktabs}
\usepackage{arydshln}
\usepackage{setspace}
\usepackage{amsthm}
\usepackage{wrapfig}
\usepackage{bm}
\usepackage{epstopdf}
\usepackage{amssymb}
\usepackage{url}
\usepackage{enumitem}
\usepackage{multirow}
\usepackage{booktabs}
\usepackage{tikz}
\usepackage{epsfig,color,amsmath,cite}
\usepackage{amsthm}
\usepackage{wrapfig}
\usepackage{bm}
\usepackage{epstopdf}
\usepackage{amssymb}
\usepackage{url}
\usepackage{enumitem}
\usepackage{multirow}
\usepackage{booktabs}
\usepackage{tikz}
\usepackage{pgfplots}
\usepackage[linesnumbered,lined,boxed,commentsnumbered,ruled,longend]{algorithm2e}



\setlist[itemize]{leftmargin=*}

\makeatother

\pgfplotsset{compat=1.16}

\allowdisplaybreaks[4]
\usepackage{graphicx}
\usepackage{color}
\usepackage{amsfonts}
\usepackage{float}
\usepackage{amsmath}
\usepackage{amsthm}
\usepackage{epstopdf}
\usepackage{changepage}
\usepackage{latexsym}
\usepackage{tikz}
\usepackage{cases}
\usepackage{longtable}
\usepackage{supertabular}
\usepackage{tabu}
\usepackage{multirow}
\usepackage{multicol}
\usepackage{booktabs}
\usepackage{arydshln}
\usepackage{setspace}

\usepackage[noabbrev]{cleveref}

\usepackage{mathtools}

\DeclarePairedDelimiter\abs{\lvert}{\rvert}%
\DeclarePairedDelimiter\norm{\lVert}{\rVert}%

\makeatletter
\let\oldabs\abs
\def\abs{\@ifstar{\oldabs}{\oldabs*}}
\let\oldnorm\norm
\def\norm{\@ifstar{\oldnorm}{\oldnorm*}}
\makeatother

\DeclareMathAlphabet\mathbfcal{OMS}{cmsy}{b}{n}

\usepackage{stackengine}

\usepackage{graphicx}
\usepackage{float}

\newcommand{\tabincell}[2]{\begin{tabular}{@{}#1@{}}#2\end{tabular}}

\usepackage{dblfloatfix} 
\usepackage[justification=centering]{caption}
\bstctlcite{IEEEexample:BSTcontrol}
\expandafter\def\expandafter\normalsize\expandafter{%
 	\normalsize
 	\setlength\abovedisplayskip{2.7pt}
 	\setlength\belowdisplayskip{2.7pt}
 	\setlength\abovedisplayshortskip{2.7pt}
 	\setlength\belowdisplayshortskip{2.7pt}
}

\begin{document}

\title{Generator Parameter Estimation by Q-Learning Based on PMU Measurements}


\author{\IEEEauthorblockN{Seyyed Rashid Khazeiynasab}
\IEEEauthorblockA{\textit{Electrical and Computer Engineering} \\
\textit{University of Central Florida}\\
Orlando, FL 32816 USA \\
rashid@knights.ucf.edu}
\and
\IEEEauthorblockN{Junjian Qi}
\IEEEauthorblockA{\textit{Electrical and Computer Engineering} \\
\textit{Stevens Institute of Technology} \\
Hoboken, NJ 07030 USA \\
jqi8@stevens.edu}
\and
\IEEEauthorblockN{Issa Batarseh}
\IEEEauthorblockA{\textit{Electrical and Computer Engineering} \\
\textit{University of Central Florida}\\
Orlando, FL 32816 USA \\
Issa.Batarseh@ucf.edu}
}

\maketitle

\begin{abstract}
In this paper, a novel Q-learning based approach is proposed for estimating the parameters of synchronous generators using PMU measurements. 
Event playback is used to generate model outputs under different parameters for training the agent in Q-learning. We assume that the exact values of some parameters in the model are not known by the agent in Q-learning. Then, an optimal history-dependent policy for the exploration-exploitation trade-off is planned. With given prior knowledge, the parameter vector can be viewed as states with a specific reward, which is a function of the fitting error compared with the measurements.  The agent takes an action (either increasing or decreasing the parameter) and the estimated parameter will move to a new state. Based on the reward function, the optimal action policy will move the parameter set to a state with the highest reward. If multiple events are available, they will be used sequentially so that the updated $\mathbfcal{Q}$-value can be utilized to improve the computational efficiency. The effectiveness of the proposed approach is validated through estimating the parameters of the dynamic model of a synchronous generator. 
\end{abstract}


\allowdisplaybreaks
\section{Introduction}

In power systems, monitoring, protection, and control are usually model-based, an accurate dynamic model for either synchronous generators \cite{nerc1996, mohiuddin2019maximum,8884738} or inverters \cite{alatrash2012generator} is thus essential.
The inaccuracy of the power system model has been witnessed in the blackout occurred in Western U.S. in 1996 \cite{nerc1996}, in which model simulations showed a stable response while the system became unstable \cite{khazeiynasab2020resilience, huang2018calibrating}. The synchronous generator is one of the most critical components in power systems and its accurate modeling is important for studying the dynamics of the system. This is no trivial task because: 1) The models may not be available for all components; 2) Even if the models are available, the parameters of the models may not be available; and 3) Even if the models and the parameters are available, the parameters may have changed over time. 

Offline methods have been proposed for estimating generator parameters, for which the synchronous machine should be out of service. Besides, these methods do not consider the changes of parameters due to aging \cite{huang2018calibrating}. 
With the increasing installation of phasor measurement units (PMUs) that provide high-quality online data to monitor the system status, there is an increasing interest in estimating synchronous generator parameters using synchrophasor data. 
The black-box optimization based method in  \cite{khazeiynasabpmu} has a good estimation accuracy, but for high-dimensional cases the estimation error will increase. Kalman filter (KF) based methods are another popular methods that have been applied to estimate the generator parameters. The extended Kalman filter (EKF) used in \cite{huang2013generator} applies a linear approximation of models and its accuracy is thus reduced \cite{wang2020calibrating}. The Ensemble Kalman filter (EnKF) \cite{huang2018calibrating} and Unscented Kalman filter (UKF) \cite{aghamolki2015identification} are further proposed to improve the accuracy of the estimation.  However, the KF based methods usually suffer from slow convergence rate \cite{surface}. 

In addition, Bayesian inference methods have been proposed for generator parameter estimation \cite{surface, xu2019bayesian}.  However, these methods need a likelihood function for implementation which may lead to computational intractability for high-dimensional cases. The accuracy of the method in \cite{surface} degrades under large parameter errors. The performance of the method in \cite{xu2019bayesian} may deteriorate for high-dimensional parameters.

Machine learning has been applied for parameter calibration. 
The method in \cite{huang2019parameters} generates extensive simulation data to train a multi-output convolutional neural network (CNN) model and predict the generator parameters. However, it requires a very large volume of data and expensive GPUs. Deep Q-learning based methods are developed in \cite{wangdrl,wang2020multi}, in which multiple events are simultaneously used for calibrating five parameters.   
Although deep Q-learning improves the scalability of Q-learning and addresses its limitation of only working for  discrete and finite state and action spaces, it is very sensitive to hyper parameters and requires a lot of tuning in order to converge. Its implementation is also much more complicated since a lot of efforts are needed to make the neural network function approximation actually work and the deep Q-learning be stabilized. For generator parameter calibration problem in which usually only a few critical parameters need to be calibrated, the benefit of deep Q-learning is not obvious.  
 
In this paper, different from \cite{wangdrl,wang2020multi}, we directly adopt Q-learning, a model-free adaptive dynamic programming algorithm, to learn the optimal policy of estimating parameters. 
The advantages and differences of our proposed approach compared with \cite{wangdrl,wang2020multi} include: 1) The implementation is much more straightforward and reliable without the many complications in deep Q-learning; 2) Q-learning works well for the parameter calibration problem with a small number of parameters to be calibrated; 3) Instead of simultaneously using multiple events we propose to use them sequentially to improve the computational efficiency; and 4) We define a different reward function based on the discrepancy between the PMU measurements and the outputs of the model. 


The remainder of this paper is organized as follows. Section \ref{sec:model} presents the dynamic model of the synchronous generator and the event playback procedure in generator model validation and parameter estimation.  Section \ref{sen} introduces the sensitivity approach for identifying critical parameters. In Section \ref{sec:qlearning}, an overview is provided for reinforcement learning (RL), especially Q-learning, and a Q-learning based method is developed for generator parameter estimation.  Section \ref{result} presents case studies to validate the proposed method. Finally concluding remarks are given in Section \ref{conclusion}.

\section{Generator Dynamic Model}\label{sec:model}

The dynamical model of a synchronous generator can be written in a general form as:
\begin{subnumcases} {\label{n1}}
\dot{\boldsymbol{x}}=\boldsymbol{f}(\boldsymbol{x},\boldsymbol{u},\boldsymbol{\alpha}) \\
{\boldsymbol{y}}=\boldsymbol{h}(\boldsymbol{x},\boldsymbol{u},\boldsymbol{\alpha}),
\end{subnumcases}
where $\boldsymbol{f}(\cdot)$ and $\boldsymbol{h}(\cdot)$ are state transition and output functions, $\boldsymbol{u} \in \mathbb{R}^v$ is the injected measured signals (voltage magnitude, phase angle,  and frequency), $\boldsymbol{y}\in \mathbb{R}^o$ is the output variables (including active and reactive power of the generator), $\boldsymbol{\alpha}\in \mathbb{R}^z$ is the parameter vector, and $\boldsymbol{x} \in \mathbb{R}^n$ is the state vector that could include rotor angle, rotor speed, transient or sub-transient voltages, and controller states.

\begin{figure*}[!t]
  \centering
  \subfloat[]{\includegraphics[width=0.24\textwidth] {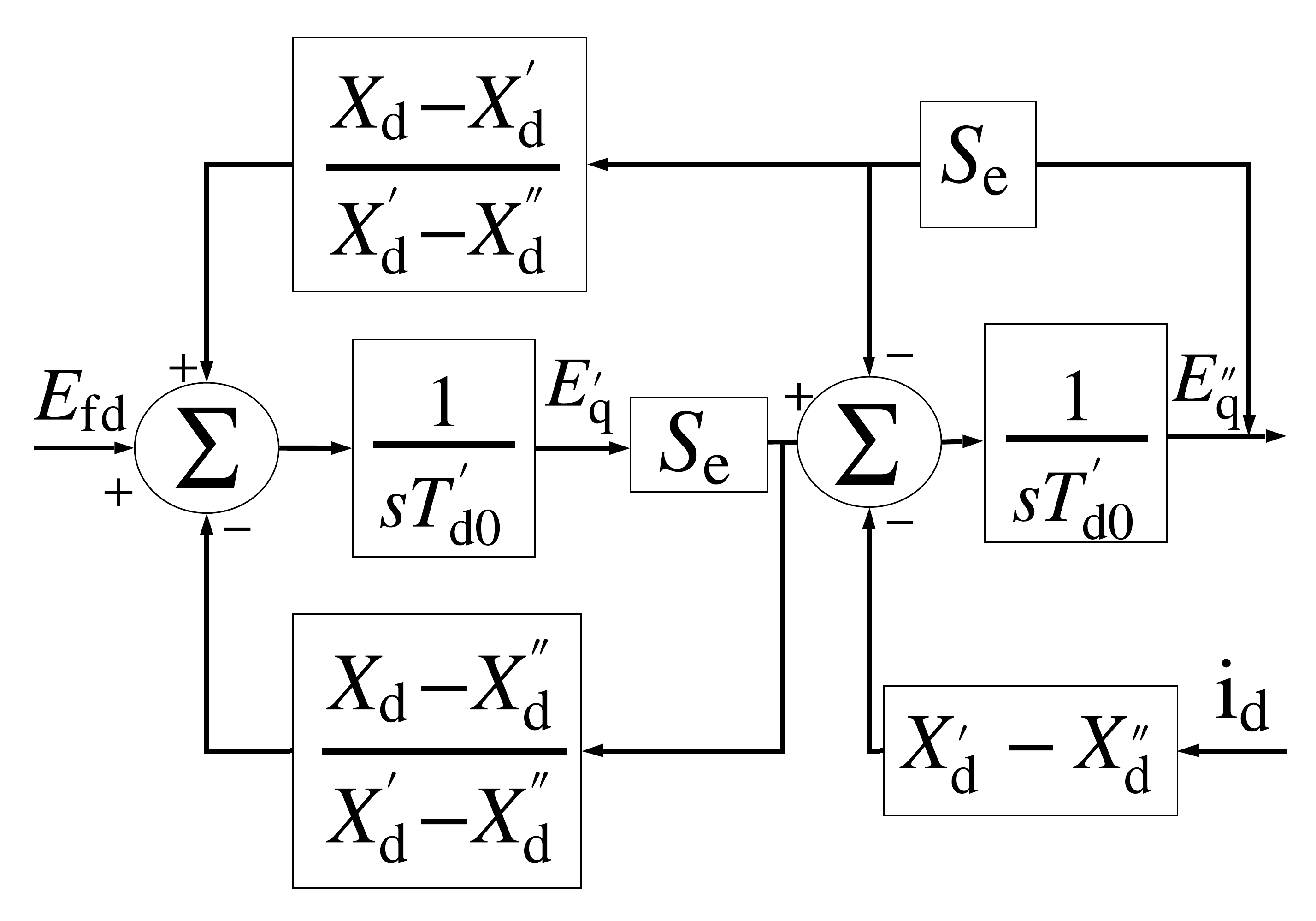}\label{gen}}
      \subfloat[]{\includegraphics[width=0.35\textwidth] {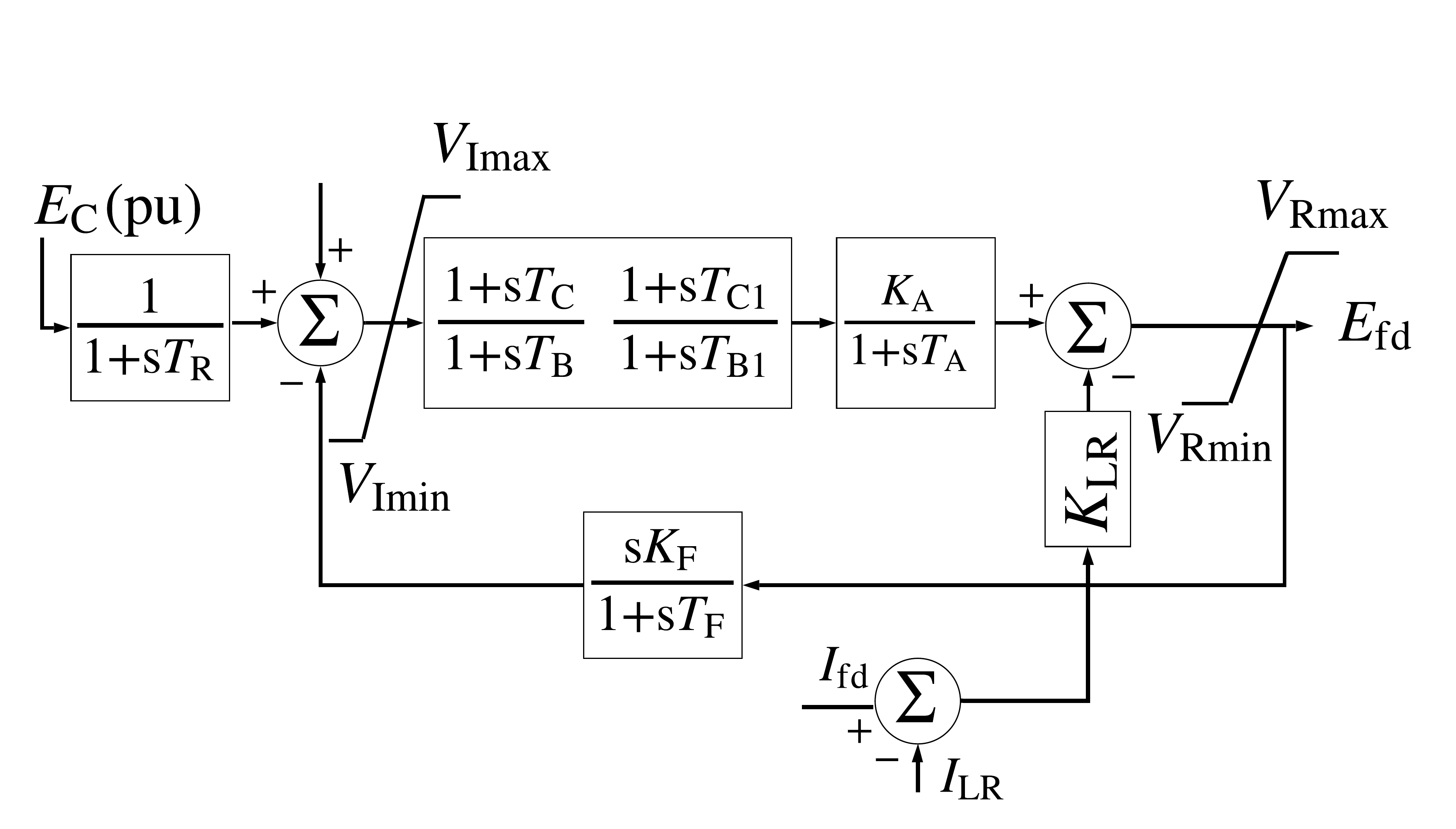}\label{exc}}
    \subfloat[]{\includegraphics[width=0.28\textwidth] {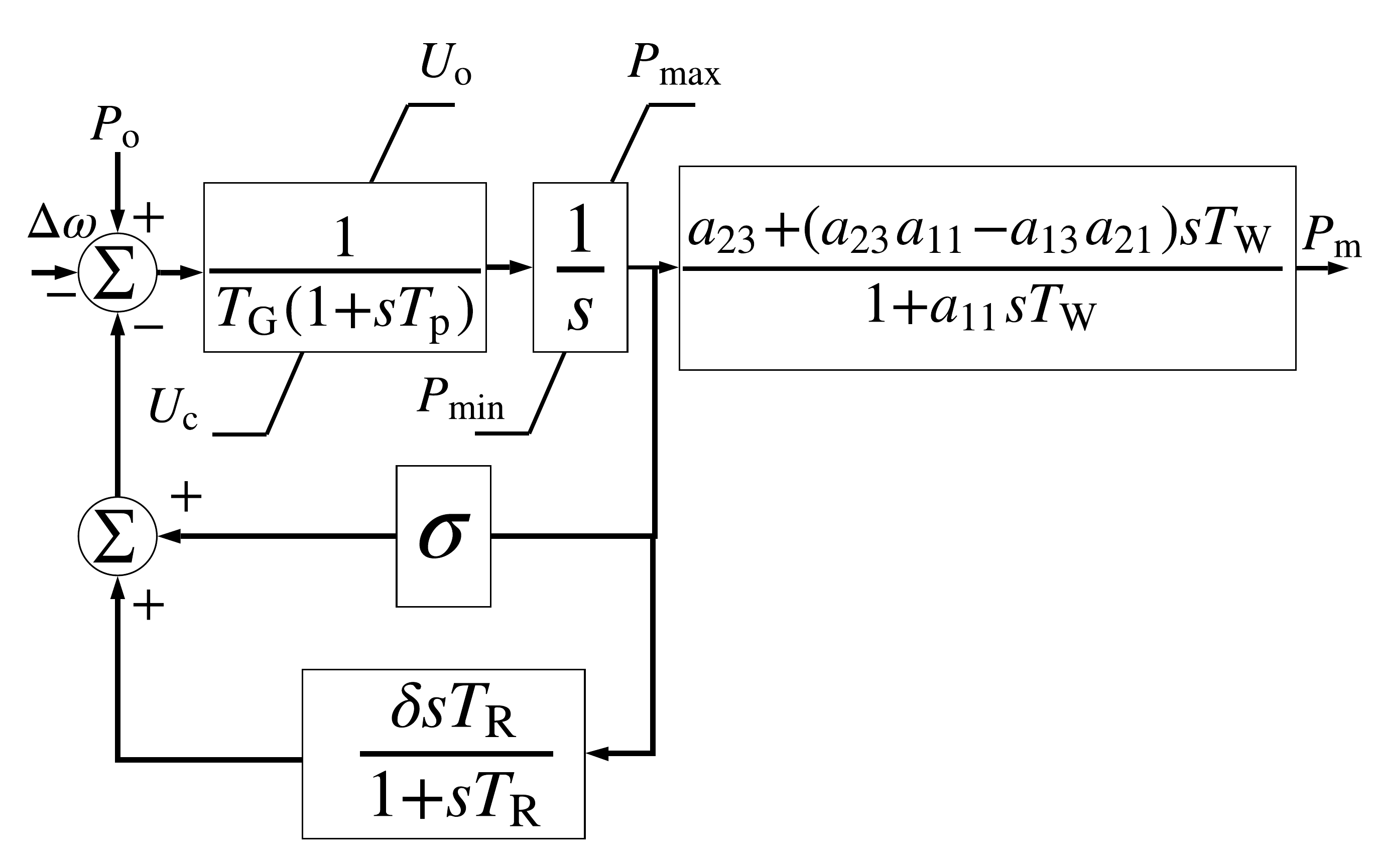}\label{gov}}
    \caption{\footnotesize Generator, exciter, and governor models \cite{siemens2009pss}: (a) d-axis GENTPJ generator model; (b) ESST1A  exciter model; (c) IEEEG3 governor model.}
  \label{models}
\end{figure*}

The model includes a synchronous machine with GENTPJ model, an exciter with ESST1A model, and a governor with IEEEG3 model. The d-axis GENTPJ model without swing equations is shown in Fig. \ref{gen}, where $X_\mathrm{d}$ and $X_\mathrm{q}$ are d- and q-axis synchronous reactance, $X'_\mathrm{d}$ and $X'_\mathrm{q}$ are d- and q-axis transient synchronous reactance, 
$X''_\mathrm{d}$ and $X''_\mathrm{q}$ are d- and q-axis sub-transient synchronous reactance, $E'_\mathrm{d}$ and $E'_\mathrm{q}$ are d- and q-axis transient voltages, $ E''_\mathrm{d}$ and $E''_\mathrm{q}$ are d- and q-axis sub-transient voltages.  $T'_\mathrm{do}$ is d-axis  transient open-circuit time constant, $T''_\mathrm{do}$ is sub-transient open-circuit time constant, $I_\mathrm{d}$ is d-axis current of the generator, and $E_\mathrm{fd}$ is the field voltage. 

The ESST1A exciter model is shown in Fig. \ref{exc}, where $K_\mathrm{A}$ and $K_\mathrm{f}$ are the main and feedback gains of the exciter, $T_\mathrm{C}$,  $T_\mathrm{B}$, $T_\mathrm{C1}$, $T_\mathrm{B1}$, and $T_\mathrm{A}$ are time constants,  $I_\mathrm{LR}$ is the maximum field current, and $K_\mathrm{LR}$ is the gain on field current of the exciter. The governor  model is shown in Fig. \ref{gov}, where $P_\mathrm{m}$ is the mechanical power, $a_{23}$, $a_{21}$, $a_{11}$, and $a_{13}$ are turbine coefficients, and the other parameters can be found in \cite{siemens2009pss}.

The ``event playback'' first uses a PMU to record the bus voltage magnitude, phase angle, frequency, and active/reactive power at the point of common coupling (PCC) during the  events.  Then by using the measurement signals it can show the mismatches between real data and the model's output \cite{huang2018calibrating, huang2013generator}.  
At time step $k$, the output of the model is \cite{huang2018calibrating}:
\begin{align}
 P_{\mathrm{model},k} &= \frac{E''_kV_k}{X'_\mathrm{d}+X_\mathrm{tr}}\sin(\delta_k-\theta_k) \label{pk} \\ 
 Q_{\mathrm{model},k} &= \frac{V^2_k-E''_kV_k\cos(\delta_k-\theta_k)}{X'_\mathrm{d}+X_\mathrm{tr}}, \label{qk}
\end{align}
where $E''_k$ and $\delta_k$ are, respectively, the generator sub-transient voltage and rotor angle, $X_\mathrm{tr}$ is the reactance of the step-up transformer, and $X'_\mathrm{d}$ is the d-axis transient synchronous reactance. 
The real and reactive power vectors from the model for all time steps, denoted by $P_{\mathrm{model}}$ and $Q_{\mathrm{model}}$, are compared with the real and reactive power measurement vectors for all time steps, denoted by $P_{\mathrm{meas}}$ and $Q_{\mathrm{meas}}$.  Let $\boldsymbol{z}^* = [P_\mathrm{meas}^\top \;  Q_\mathrm{meas}^\top]^\top$ be the measurement from PMU and $\boldsymbol{z} = [P_\mathrm{model}^\top \;  Q_\mathrm{model}^\top]^\top$ be the output of the model.

\section{Identifying Critical Parameters} \label{sen}

After a model deficiency is revealed, the next step is to identify the problematic parameters. A generator with its control can have many parameters. Calibrating all of them could be computationally challenging and also not every parameter is identifiable. Trajectory sensitivity has been used to identify the most critical parameters \cite{huang2013generator}. The sensitivity of the output with regard to parameter $\alpha_i$ can be calculated as \cite{huang2018calibrating}: 
\begin{align}
S(\alpha_i) &=\frac{1}{2K}\sum_{k=1}^{2K}{\frac{\alpha_i|z_k(\alpha_i^+)-z_k(\alpha_i^-)| }{\alpha_i^+-\alpha_i^-}}, \notag
\end{align}
where 
$K$ is the number of time steps, $\alpha_i^+=\alpha_i + \Delta \alpha_i$ and $\alpha_i^-=\alpha_i-\Delta \alpha_i$, and $\Delta \alpha_i$ is a small perturbation of $\alpha_i$. 
After sensitivity analysis, the parameters selected to be estimated are those with large sensitivities \cite{huang2013generator, huang2018calibrating}. 

\section{Q-Learning for Generator Parameter Estimation}\label{sec:qlearning}

Q-learning is a model-free RL algorithm with the goal of learning a policy to tell an agent what action to take under what circumstances. In Q-learning, an agent takes sequential actions at a series of states based on a state-action value matrix,
Q-table, until reaching an ultimate goal \cite{watkins1989learning}.
Let   $\mathcal{A}$ and $\mathcal{S}$ be
the action space and state space respectively. At each episode $t$, the agent observes a state $ s_t \in \mathcal{S}$ and chooses an action $a_t \in \mathcal{A}$  based on policy $\pi$, which is a function that maps states into actions. As a consequence of taking  action $a_t$, the agent receives a reward $R_t$ defined as $R_t = R(s_t, a_t, s_{t+1})$ and observes the next state $s_{t+1}$ of the environment. The RL framework considers the Markov decision process assumption, i.e $s_{t+1}$ is only conditioned by  $s_t$ and $a_t$ and is sampled according to the transition probability $p(s_{t+1}\vert s_t, a_t )$. The above process is continued until the agent reaches the last episode, called the terminal state \cite{seo2019rewards}.

The goal of the agent is to take actions so as to maximize the expected return for a given state $s_t$. The expected return for selecting action $a$ in state $s$, action-value function following a policy $\pi$, is defined as $\mathbfcal{Q}^\pi(s,a) = \mathbb{E}[R_t\vert s=s_t,a = a_t , \pi ]$ \cite{seo2019rewards}. 
The expected return at episode $t$ when the action $\pi^*$ maximizes the $\mathbfcal{Q}$ function 
is given as: 
\begin{align}
\mathbfcal{Q}^\pi(s_t,a_t) = \mathbb{E}\big[R_t+ \gamma\,  \mathrm{max}\underset{a_{t+1} \in \mathcal{A}}{\mathbfcal{Q}^*}(s_{t+1}, a_{t+1})\big],
\end{align}
where $\gamma \in [0 ,1]$ is the discount factor that weights the future rewards. The state-action value functions are updated by \cite{sutton2018reinforcement}:
\begin{align}\label{eq:Q}
&\mathbfcal{Q}_{t+1}(s_{t+1}, a_{t+1}) = (1-\lambda) \mathbfcal{Q}_t(s_t,  a_t)\notag \\
&+ \lambda\big[R(s_t, a_t, s_{t+1})+ \gamma\,  \mathrm{max}\underset{a_{t+1}\in A}{\mathbfcal{Q}_t}(s_{t+1}, a_{t+1})\big], 
\end{align}
where $\lambda$ is the learning factor that controls the aggressiveness of the learning. The balance of exploration and exploitation in Q-learning is maintained by adopting a
decaying $\varepsilon_{\mathrm{t}}$-greedy method \cite{seo2019rewards, sutton2018reinforcement}, by which the agent takes random actions at the beginning while reducing the randomness during the learning process \cite{sutton2018reinforcement}. The optimal policy at episode $t$ is represented by a greedy strategy as:
\begin{align}
    a_t = \underset{a \in A}{\arg\max} \; \mathbfcal{Q}_{t}(s, a).
\end{align}
The values of $\mathbfcal{Q}$ are initialized to be zero and are updated repeatedly by (\ref{eq:Q}) based on the action reward in the current state and
the maximum reward in the next state. The algorithm will converge to the optimal policy, $\mathbfcal{Q}^*$, after $N$ episodes.

\subsection{Design of State and Action}

The state is the coordinate information of the spots that the agent moves in the parameter space:
\begin{align}
    & \mathcal{S} =  \{s_1, s_2, \cdots, s_{N_\mathrm{s}} \},
\end{align}
where $N_\mathrm{s}$ is the number of states, which equals the number of select-able states in the parameter space. 
For each element $1\leq j\leq L$ in the parameter vector $\boldsymbol{\alpha}_{\mathrm{c}} = [\alpha_{\mathrm{c}, 1},\cdots ,\alpha_{\mathrm{c}, L}]^\top $, if the upper bounds and lower bounds of their priors are $\boldsymbol{\alpha}_{\mathrm{c}}^{\mathrm{u}} = [\alpha_{\mathrm{c},1}^\mathrm{u},\cdots ,\alpha_{\mathrm{c}, L}^\mathrm{u}]^\top $ and $\boldsymbol{\alpha}_{\mathrm{c}}^\mathrm{l} = [\alpha_{\mathrm{c}, 1}^\mathrm{l},\cdots ,\alpha_{\mathrm{c}, L}^\mathrm{l}]^\top $respectively. We define a maximum acceptable error as $\tau$, then we discrete the domain for parameter $j$ by $\Delta s_j$ given as: 
\begin{align}\label{deltasj}
   \Delta s_j = 2\tau\big(\alpha_{\mathrm{c},j}^\mathrm{u}-\alpha_{\mathrm{c},j}^\mathrm{l}\big).
\end{align}
For a parameter with initial value 5 and acceptable error as $1\%$, if the prior is uniform as $\mathcal{U}(0,10)$, then 
$\Delta s = 0.2$ and the number of states for this parameter is equal to 50. 

The action space $\mathcal{A}$ is composed of the select-able movement in the parameter space: 
\begin{align}
    & \mathcal{A} =  \{a_1, a_2, \cdots, a_{N_\mathrm{a}} \},
\end{align}
where $N_\mathrm{a}$ is the number of actions. The actions are increasing or decreasing the parameter $j=1,\ldots,L$ by $\Delta s_j$ in the states. 
For example, when there are two parameters to be calibrated, we have four different actions: up, down, right, and left. The new state is updated according to the chosen action.

\subsection{Design of Dynamic Reward Function}

For some RL problems, the reinforcement signal may not appear immediately after the action that triggers the rewards \cite{seo2019rewards}. In this type of problems, RL faces delayed rewards. Handling delayed rewards is one of the major challenges in RL. The agent must interact with the environment to
adapt to it, and may need to spend a lot of time in attaining the optimal behavior. Reward functions describe how the agent ``ought" to behave in this process based on which the agent learns how to move in the environment \cite{seo2019rewards}. To accelerate this learning process, the reward function should describe the agent’s state in a timely and accurate manner. If the reward function is better behaved, the agent will learn better. Therefore, the design of the reward function is critical for RL. 

True parameters are assumed to be unknown. The reward function is based on the discrepancy between the simulated outputs and the measurements,  $\epsilon_\mathrm{s}(\boldsymbol{z},\boldsymbol{z}^*)$, is defined below:    
\begin{align} \label{distance_function}
&\epsilon_\mathrm{s}(\boldsymbol{z},\boldsymbol{z}^*) = \frac{1}{2K}\norm{\boldsymbol{z}-\boldsymbol{z}^*}_1,
\end{align} 
where $\norm{\cdot}_1$ is the 1-norm of a vector. Based on the predefined thresholds $\bar{\epsilon}$ and $\underline{\epsilon}$, the reward function is defined as: 
\begin{align}\label{reward_function}
R(s,a)&= 
\begin{cases}
\frac{10}{\epsilon_\mathrm{s}+0.01}, &\text{if $\, \epsilon_\mathrm{s} < \underline{\epsilon}$} \\
0, &\text{if $\, \underline{\epsilon} \le \epsilon_\mathrm{s} \le \bar{\epsilon}$} \\
-10(\epsilon_\mathrm{s}-\bar{\epsilon}), &\text{if $\, \epsilon_\mathrm{s} > \bar{\epsilon}$}. 
 \end{cases} 
 \end{align}
If $\epsilon_\mathrm{s}<\underline{\epsilon}$, the agent gets positive rewards and the states with smaller $\epsilon_\mathrm{s}$ get larger rewards;  
if $\underline{\epsilon} \le \epsilon_\mathrm{s} \le \bar{\epsilon}$, the agent does not get any rewards; if $\epsilon_\mathrm{s}>\bar{\epsilon}$, the agent gets negative rewards.  

\subsection{Proposed Algorithm for Generator Parameter Estimation}

We define a matrix $\mathcal{O}$ to show the status of every state during the exploration and exploitation. Each state has a `searched' or `un-searched' status, and when the agent visits a state, the status of that state is set to be `searched'. This matrix is useful for guaranteeing to run the model in each state only once. If a state is explored by the agent for the first time, we calculate $\epsilon_\mathrm{s}(\boldsymbol{z},\boldsymbol{z}^*)$ and assign a reward for it based on  (\ref{reward_function}). 
The proposed Q-learning based algorithm for generator parameter estimation is presented in Algorithm \ref{alg:Q} \cite{watkins1989learning}. 
In this algorithm $\mathrm{Model}(\boldsymbol{\alpha}_\mathrm{c}^*)$ generates the model outputs $\boldsymbol{z}$ under parameter $\boldsymbol{\alpha}_\mathrm{c}^*$ using the generator model and the event playback procedure in Section \ref{sec:model}. If multiple events are available, we will use them sequentially so that the updated $\mathbfcal{Q}$-value obtained from the training for the previous events can be utilized to significantly improve the computational efficiency.

\begin{algorithm}[!t]
\captionsetup{labelfont={sc,bf}, labelsep=newline}
  \caption{Q-Learning based method for estimating the parameter $\boldsymbol{\alpha}_\mathrm{c}$ }
\begin{algorithmic}[1]
\STATE Set hyper parameters $\lambda, \gamma, \varepsilon, N$
\STATE Initialize experience pool $\mathbfcal{O}$ as an empty set
\STATE Discretize the parameter space based on (\ref{deltasj})
\STATE $\mathbfcal{Q} \leftarrow{\boldsymbol{0}}$
\FOR{$1\leq t \leq N$}
\STATE Start with an un-searched state with parameter $\boldsymbol{\alpha}_\mathrm{c}^*$
\STATE Generate data $\boldsymbol{z}$ from $\boldsymbol{\alpha}_\mathrm{c}^*$: $\boldsymbol{z}\sim \mathrm{Model}(\boldsymbol{\alpha}_\mathrm{c}^*)$
\STATE Calculate discrepancy $\epsilon_\mathrm{s}(\boldsymbol{z},\boldsymbol{z}^*)$ based on (\ref{distance_function})
\STATE Assign a reward to the state based on (\ref{reward_function}) or $\mathbfcal{O}$
\STATE With probability $\varepsilon$, select a random action $a_t$;
otherwise select $a_t = \underset{a \in A}{\arg\max} \; \mathbfcal{Q}_{t}(s, a)$
\STATE Update $\mathbfcal{Q}$-value matrix based on (\ref{eq:Q})
\STATE Update $\mathbfcal{O}$-value matrix based on (\ref{reward_function})
\ENDFOR 
\end{algorithmic}\label{alg:Q}
\end{algorithm}

\section{Simulation Results} \label{result}

We implement our method based on PSS/E and Python 2.7 and test it on the same system used in \cite{ppmv}. 
All tests are carried out on a PC with Intel(R) Core(TM) i7-8700 and 8 GB RAM. 
A PMU is installed at the 230-kV level of the substation. The sampling rate of the PMU is 30 sample/s.

\subsection{Hyperparameter Setting}

The discount factor $0<\gamma<1$, which makes a trade-off between the immediate and long-term reward. 
In this paper, we consider $\gamma = 0.9$. 
The learning rate $\lambda$ determines the learning rate of the agent when updating $\mathbfcal{Q}$-value of each state-action pair. We choose this parameter to be $0.3$. The number of episodes $N$ is chosen as 2000. The $\varepsilon_{\mathrm{t}}$ that determines the balance of exploration and exploitation in Q-learning is chosen as $0.2$ \cite{seo2019rewards}. We choose $\underline{\epsilon} = 0.001$ and $\bar{\epsilon}=2$ in (\ref{reward_function}).

\subsection{Critical Parameter Identification}

For sensitivity analysis, a small perturbation $\Delta \alpha_i=5\%|\alpha_i|$ is applied to each parameter.
The top four parameters and their sensitivities are listed in Table \ref{sensitivity}. These are the gain of the exciter $K_\mathrm{A}$, the time constant of the exciter ($T_\mathrm{b}$), the turbine coefficient ($a_{23}$), and the d-axis  transient  rotor  time  constant ($T_\mathrm{pdo}$). They are identified as critical parameters.

\begin{table}[!t]
\centering
\footnotesize
\renewcommand{\arraystretch}{0.8}
\captionsetup{labelsep=space,font={footnotesize,sc}}
\caption{\\Sensitivity Analysis of Parameters}
\begin{tabular}{c|c}
\hline\hline 
\textbf{Parameter} & \textbf{Sensitivity} \\ \hline
      $K_\mathrm{A}$   &   1.75          \\ \hline
        $T_\mathrm{B}$  &    1.35         \\ \hline
         $a_{23}$ &   1.13          \\ \hline
        $T_\mathrm{pdo}$  &     1.11        \\ \hline\hline
\end{tabular}\label{sensitivity}
\end{table}

\subsection{Estimation for Two Parameters} \label{two_para_case}
  
In practical implementation, the parameters provided by the manufacturer may change due to aging. To consider this uncertainty, we assume that the available parameters are deviated from the true values. We estimate the same two parameters as in \cite{xu2019bayesian}, which are the moment of inertia $H$ for synchronous generator and the amplifier gain $K_\mathrm{A}$ for the exciter. Their true values are $H_\mathrm{True} = 5.4$ and $K_{\mathrm{A}_\mathrm{True}} = 125$. We assume that the mean values of these parameters are 10\% greater than their true values to account for parameter uncertainties. We choose the lower/upper bounds of the uniform prior distributions as 50\% or 70\% less/greater than the mean values, respectively. The results of  parameter estimation under different prior distributions are shown in Table \ref{diffp}. It is seen that the proposed method can provide accurate estimation of the parameters.

\begin{table}[!t]
\centering
\footnotesize
\renewcommand{\arraystretch}{0.85}
\captionsetup{labelsep=space,font={footnotesize,sc}}
\caption{\\Parameter Calibration Under Different Prior Distributions}
\begin{tabular}{c|c|c|c}
\hline\hline
\multicolumn{2}{cV{1}}{ \tabincell{c}{\bf{$H$} ($H_\mathrm{True}=5.40$)}} & \multicolumn{2}{c}{  \tabincell{c}{\bf{$K_\mathrm{A}$}  ($K_{\mathrm{A}_\mathrm{True}}=125$)}} 
\\ \hline  
\textbf{Prior} & \tabincell{c}{\textbf{Estimated} \\ (Error (\%))}  & \textbf{Prior} & \tabincell{c}{\textbf{Estimated} \\ (Error (\%))} \\ \hline 
$\mathcal{U}$(2.9,\, 8.9) & \tabincell{c}{5.39 \\ (0.1)}  & $\mathcal{U}$(68.8, 206.3) & \tabincell{c}{125.8\\ (0.6)}  \\ \hline
\textbf{$\mathcal{U}$(1.8,   10.1)} & \textbf{\tabincell{c}{5.37 \\ (0.6)}}  & \textbf{$\mathcal{U}$(41.2,  233.8)} & \textbf{\tabincell{c}{123.8 \\ (0.9)}}  \\ \hline\hline
\end{tabular} \label{diffp}
\end{table}

Fig. \ref{rew-two} shows the cumulative rewards for the case in the last row of Table \ref{diffp}. It is seen that the training converges after 421 iterations, which takes 5 hours.

\subsection{Estimation for Four Parameters}

\begin{figure}[!t]
  \centering
  \subfloat[]{\includegraphics[height=1.15in]{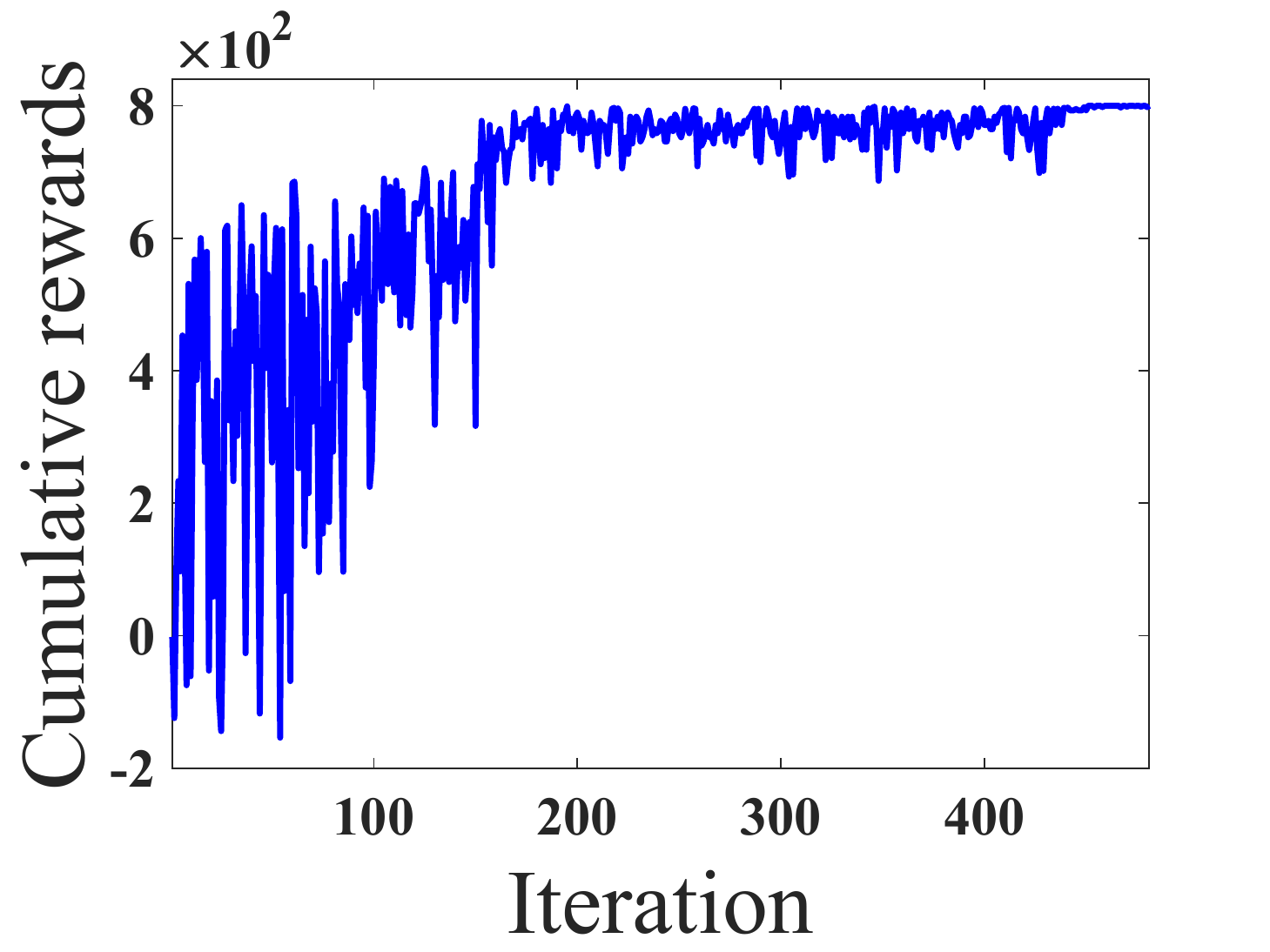}\label{rew-two}}
  \subfloat[]{\includegraphics[height=1.15in]{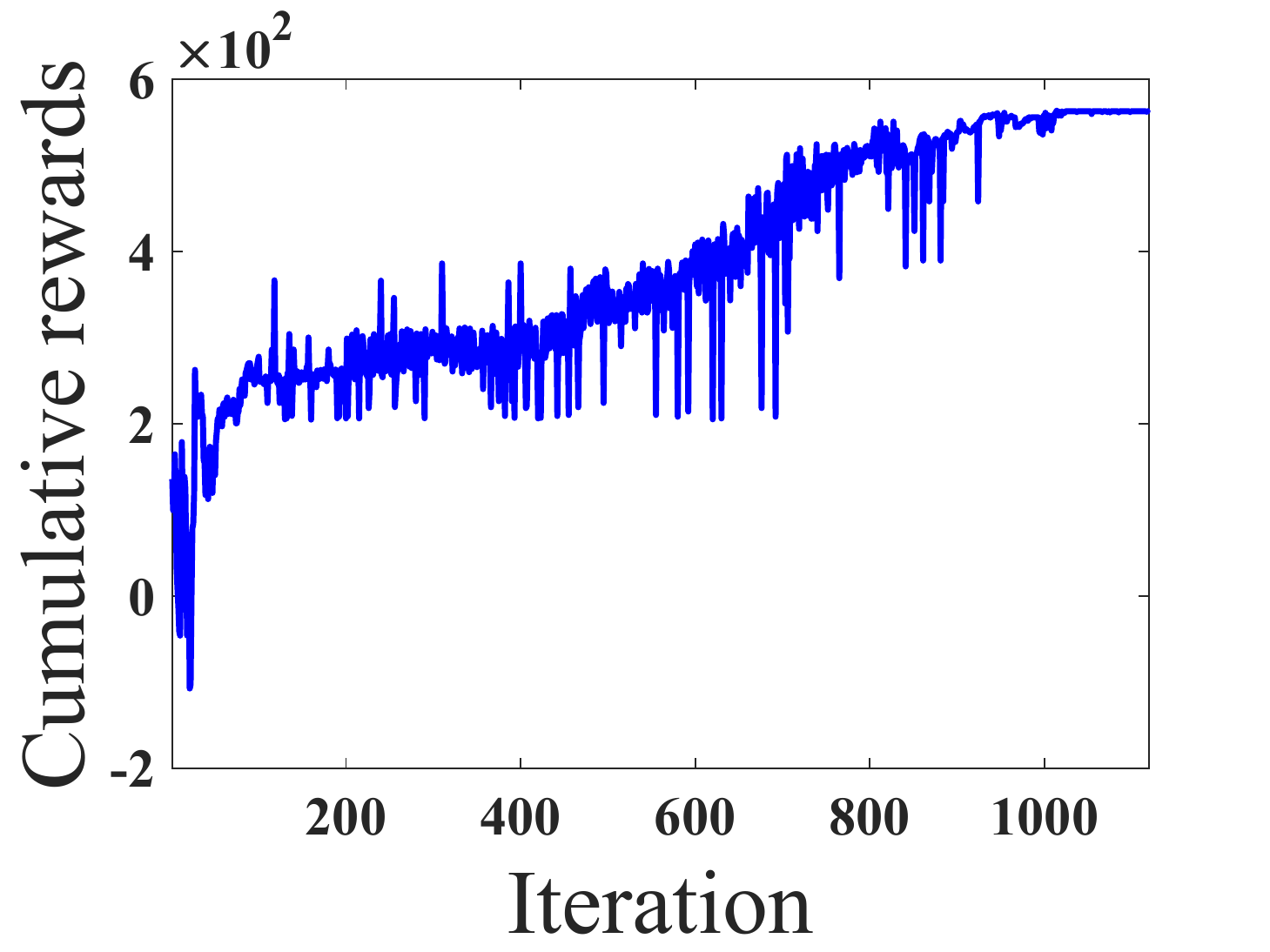}\label{rew-four}}
      \captionsetup{justification=raggedright,singlelinecheck=false}
  \caption{\footnotesize Cumulative rewards for two and four-parameter case: (a) Two parameter case; (b) Four parameter case. } 
  \label{rewards}
\end{figure}

\begin{figure}[!t]
  \centering
  \subfloat[]{\includegraphics[height=1.15in]{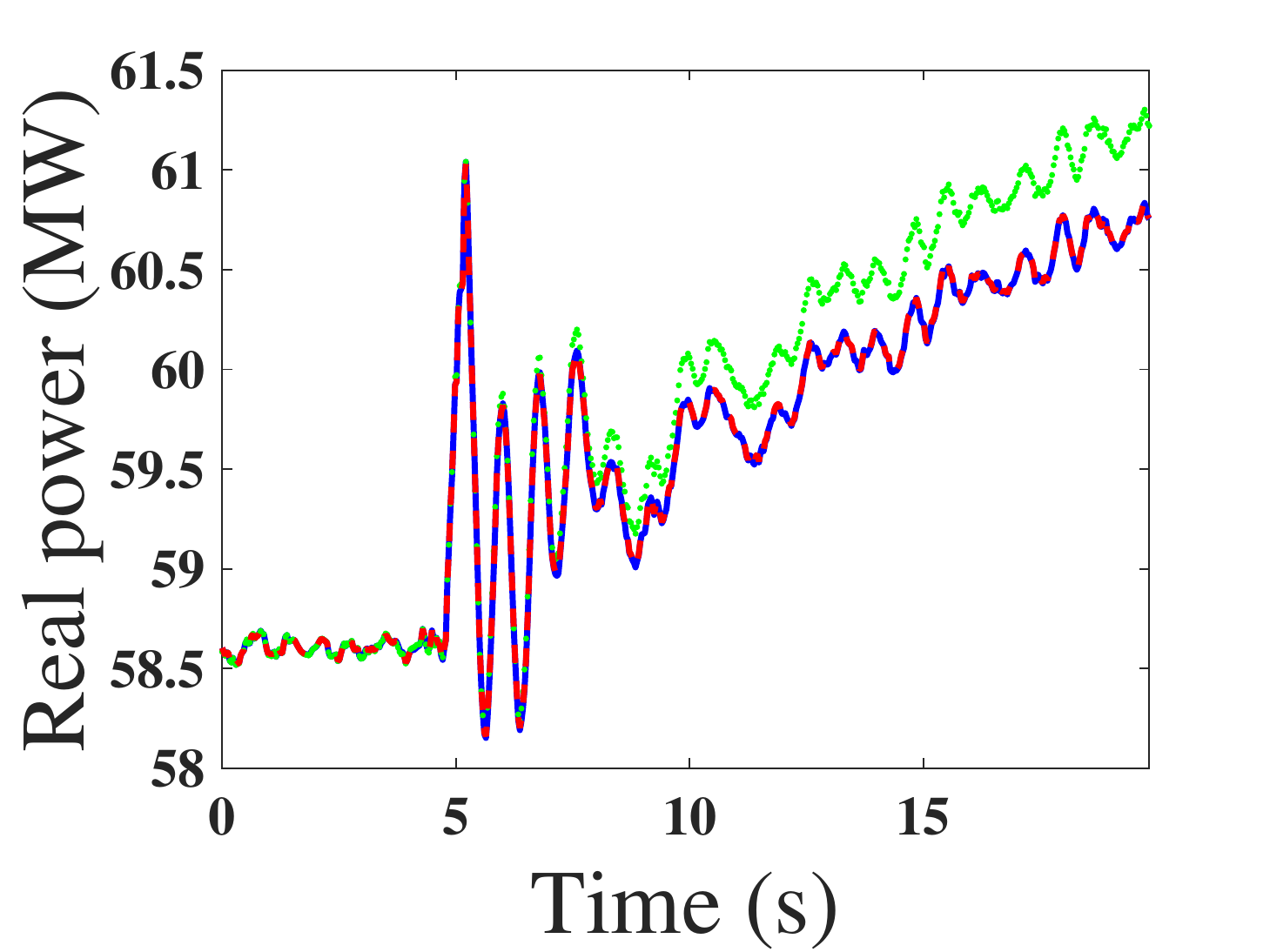}\label{real_power}}
  \subfloat[]{\includegraphics[height=1.15in]{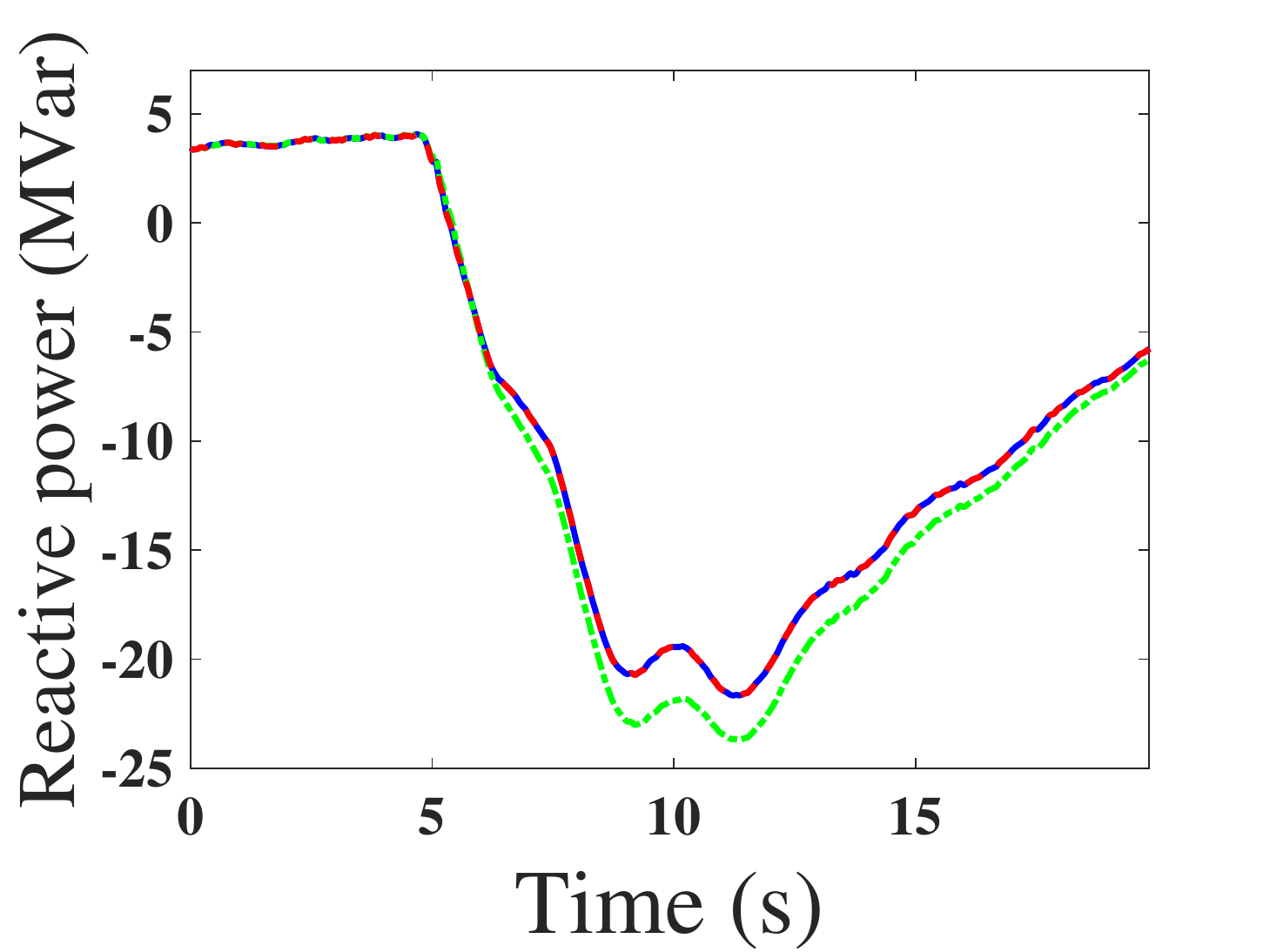}\label{reactive_power}}
      \captionsetup{justification=raggedright,singlelinecheck=false}
  \caption{\footnotesize Model performance before and after parameter calibration: (a) Real power; (b) Reactive power. \textcolor{blue}{\textbf{\textemdash}} PMU measurements; \textcolor{green}{\textbf{.}}  model before calibration; \textcolor{red}{\textbf{- -}}  model after calibration. } 
  \label{P_Q}
\end{figure}

We also demonstrate the performance of the proposed method for estimating the four critical parameters identified in Table \ref{sensitivity}. 
We assumed that the mean values of these parameters are 10\% greater than their true values. 
We choose the lower/upper bound of the uniform prior distributions as 50\% less/greater than the mean values. 
Table \ref{fourteen} lists the estimated values and the percentage errors. It is seen that the estimate is very close to the true values. 
Fig. \ref{rew-four} shows the corresponding cumulative rewards. The training converges after 1000 iterations which takes 8 hours. 
Fig. \ref{P_Q} shows the results for real and reactive power under the estimated parameters and the parameters before the calibration. We consider the parameters before the calibration as 10\% greater than the true values. Before calibration there is obvious discrepancy between the model outputs and the PMU measurements while with the estimated parameters the model outputs match the PMU measurements very well.

\begin{table}[!t]
  \footnotesize
\centering
\renewcommand{\arraystretch}{0.85}
\captionsetup{labelsep=space,font={footnotesize,sc}}
\caption{\\Calibration of Four Critical Parameters}
\label{fourteen}
\begin{tabular}{c|c|c|c}
\hline\hline 
\textbf{Parameter} & \textbf{True value} & \textbf{Estimated value} &\textbf{Error (\%)}  \\ \hline
$K_\mathrm{A}$  & 125  & 123.8 & 0.9\\ \hline
$T_\mathrm{b}$ & 3.86  &  3.82& 1\\\hline
$a_{23}$ & 1.102  & 1.10&0.1 \\ \hline
$T_\mathrm{pdo}$ & 5.4  & 5.34 & 1\\ \hline\hline
\end{tabular}
\end{table}

\subsection{Offline Training and Online Application}

\begin{figure}[!t]
  \centering
  \subfloat[]{\includegraphics[height=1.15in]{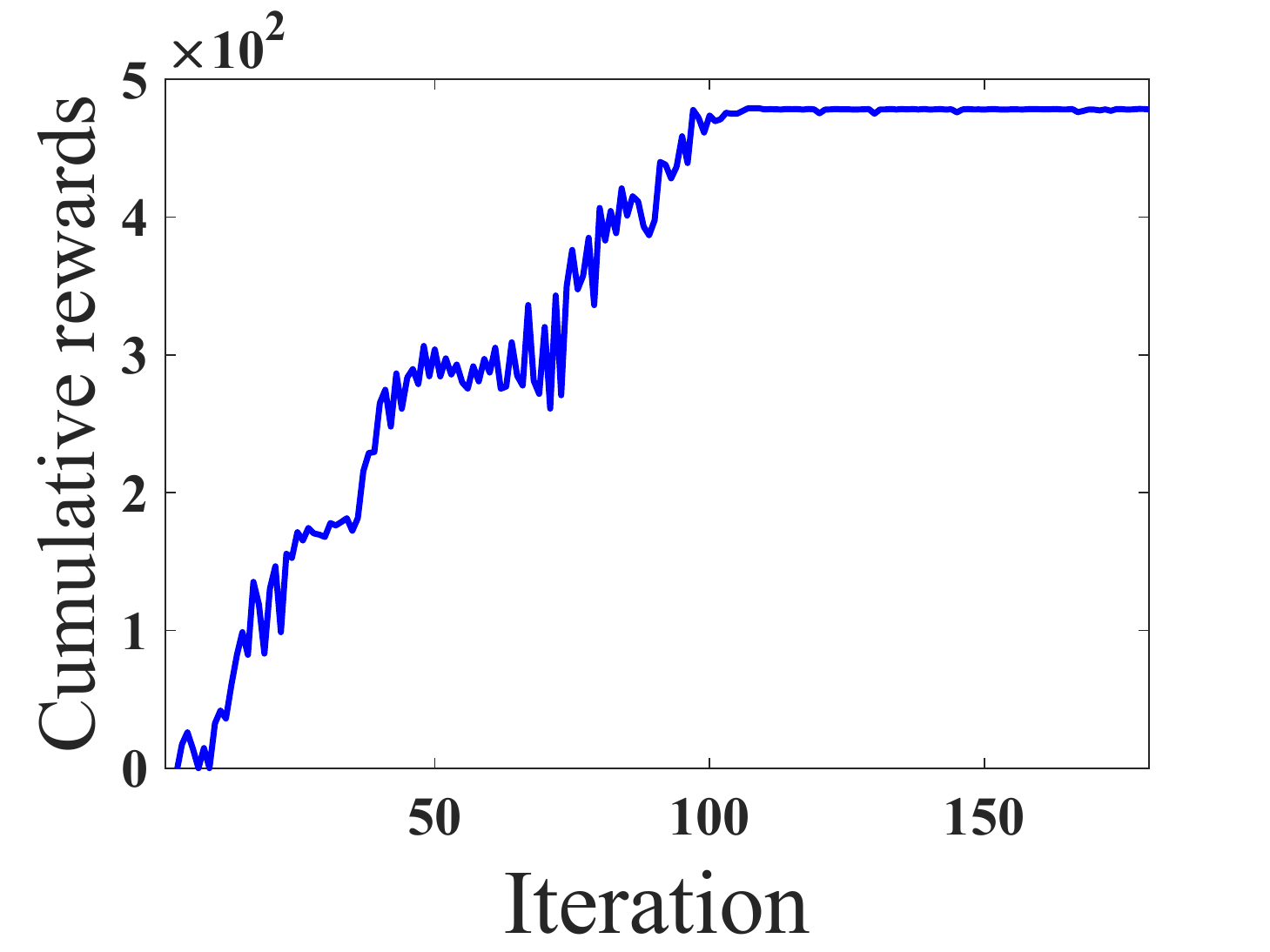}\label{secondev}}
  \subfloat[]{\includegraphics[height=1.15in]{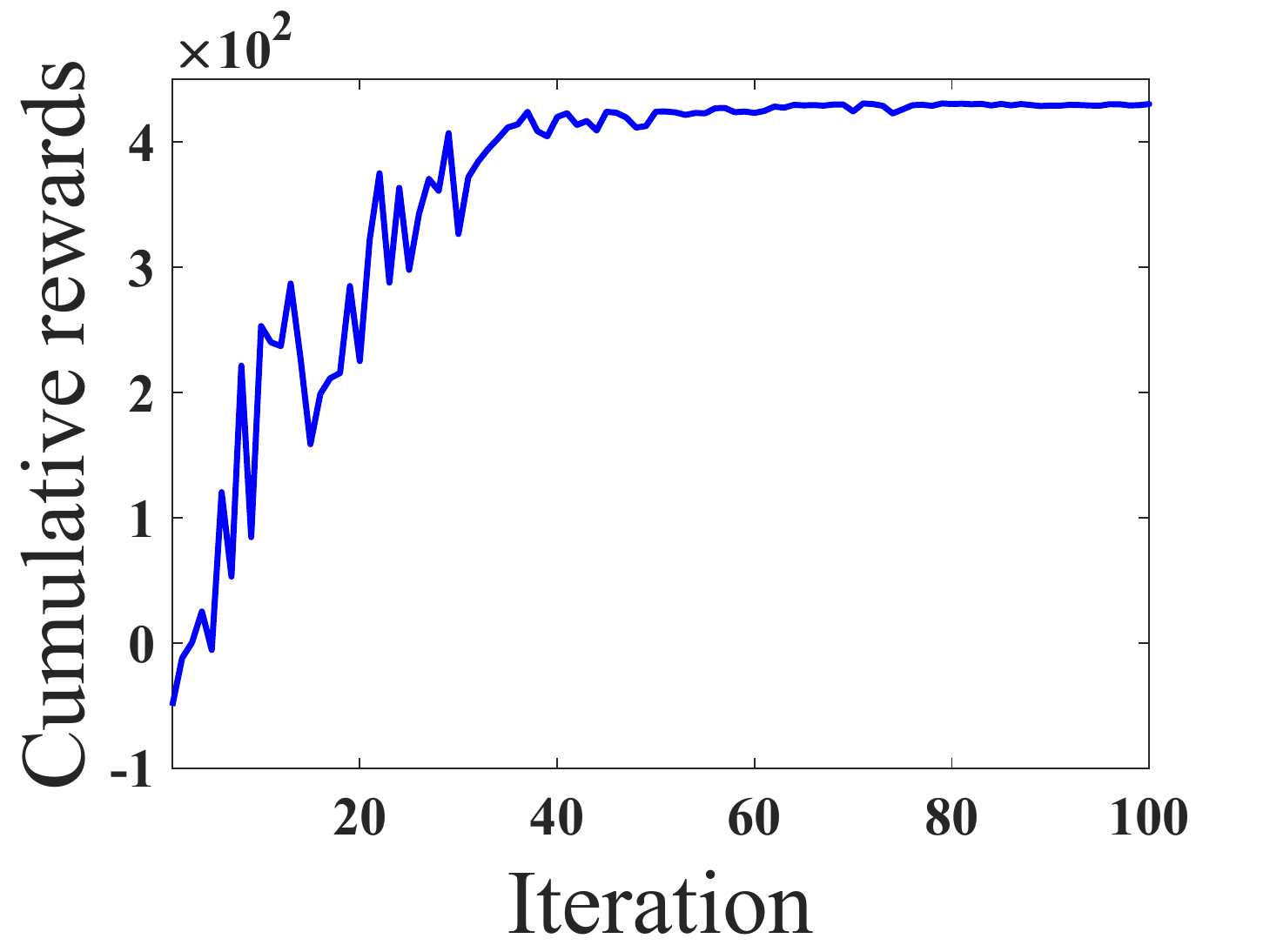}\label{secondtru}}
      \captionsetup{justification=raggedright,singlelinecheck=false}
  \caption{\footnotesize Cumulative rewards for (a) another event and (b) a case with different true parameters. } 
  \label{mevent}
\end{figure}

Parameter estimation is also conducted for another event and for the case in which the true parameters are different from those in the case in Section \ref{two_para_case}. For these cases, we use the updated $\mathbfcal{Q}$-value in Section \ref{two_para_case} which can be considered as offline training. The cumulative reward for the second event is shown in Fig. \ref{secondev}. It can be seen that the agent learns to find the true parameters in 110 iterations which is much less than that needed for the first event. 
For the case in which the true parameters are different, we use the outputs of the model under $H=4.4$ and $K_\mathrm{A}=100$ as PMU measurements and use the same prior distributions as in Section \ref{two_para_case}. As can be seen from Fig. \ref{secondtru}, the agent learns to find the true parameter after 50 iterations which is less than the number of iterations in the case in Section \ref{two_para_case}. The estimations using updated $\mathbfcal{Q}$-value take around 5 minutes. The results here show that using the updated $\mathbfcal{Q}$-value obtained from offline training can significantly improve the computational efficiency of the online parameter estimation. This also indicates that the computational efficiency can be greatly improved if multiple events are used sequentially so that the learning for a later event can leverage the updated $\mathbfcal{Q}$-value obtained from the learning for a previous event.

\section{Conclusion} \label{conclusion}

This paper proposes a Q-learning based method for generator model parameter estimation using PMU measurements. 
The simulation results for a decentralized generator show that the proposed method can accurately estimate the system parameters and the updated $\mathbfcal{Q}$-value obtained from offline training can significantly improve the computational efficiency of the online parameter estimation. 
Improving the scalability of Q-learning and addressing curse of dimensionality while maintaining its relatively simple implementation is  a very challenging problem and will be studied in our future research.  
\vspace{-0.4cm}
\bibliographystyle{IEEEtran}	
\bibliography{Main}

\end{document}